\documentclass[iop, tighten, numberedappendix, letterpaper]{emulateapj}

\usepackage{verbatim}
\usepackage{longtable}
\usepackage{amsmath}

\def\h2{$\rm H_2$}

\newcommand{\msun}{M$_{\odot}$}
\newcommand{\halpha}{H$\alpha$}

\newcommand{\hii}{H{\sc II}}

\begin{document}

\shortauthors{Weisz et al.}

\title{Modeling the Effects of Star Formation Histories on \halpha\ and Ultra-violet Fluxes in Nearby Dwarf Galaxies\footnote{Based on observations made with the NASA/ESA Hubble Space Telescope, obtained from the Data Archive at the Space Telescope Science Institute, which is operated by the Association of Universities for Research in Astronomy, Inc., under NASA contract NAS 5-26555}}

\author{Daniel R. Weisz\altaffilmark{1,2},
Benjamin D.\ Johnson\altaffilmark{3},
L. Clifton Johnson\altaffilmark{1}, 
Evan D.\ Skillman\altaffilmark{2},
Janice C.\ Lee\altaffilmark{4, 9},  
Robert C.\ Kennicutt, Jr.\altaffilmark{5}, 
Daniela Calzetti\altaffilmark{6},
Liese van Zee\altaffilmark{7}, 
Matthew S.\ Bothwell\altaffilmark{5}, 
Julianne J.\ Dalcanton\altaffilmark{1},
Daniel A.\ Dale\altaffilmark{8},
Benjamin F.\ Williams\altaffilmark{1}
}

\altaffiltext{1}{University of Washington; dweisz@astro.washington.edu}
\altaffiltext{2}{University of Minnesota}
\altaffiltext{3}{Institut d'Astrophysique de Paris}
\altaffiltext{4}{Observatories of the Carnegie Institution of Washington}
\altaffiltext{5}{University of Cambridge}
\altaffiltext{6}{University of Massachusetts}
\altaffiltext{7}{Indiana University}
\altaffiltext{8}{University of Wyoming}
\altaffiltext{9}{Carnegie Fellow}

\begin{abstract}

We consider the effects of non-constant star formation histories (SFHs) on \halpha\ and GALEX far ultra-violet (FUV) star formation rate (SFR) indicators.  Under the assumption of a fully populated Chabrier IMF, we compare the distribution of \halpha-to-FUV flux ratios from $\sim$ 1500 simple, periodic model SFHs with observations of 185 galaxies from the \emph{Spitzer} Local Volume Legacy survey.  We find a set of SFH models that are well matched to the data, such that more massive galaxies are best characterized by nearly constant SFHs, while low mass systems experience bursts amplitudes of $\sim$ 30 (i.e., an increase in the SFR by a factor of 30 over the SFR during the inter-burst period), burst durations of tens of Myr, and periods of $\sim$ 250 Myr; these SFHs are broadly consistent with the increased stochastic star formation expected in systems with lower SFRs.  We analyze the predicted temporal evolution of galaxy stellar mass, $R$-band surface brightness, \halpha-derived SFR, and blue luminosity, and find that they provide a reasonable match to observed flux distributions.  We find that our model SFHs are generally able to reproduce both the observed systematic decline and increased scatter in \halpha-to-FUV ratios toward low mass systems, without invoking other physical mechanisms.  We also compare our predictions with those from the Integrated Galactic IMF theory with a constant SFR. We find that while both predict a systematic decline in the observed ratios, only the time variable SFH models are capable of producing the observed population of low mass galaxies ($M_{*}$ $\lesssim$ 10$^{7}$ \msun) with normal \halpha-to-FUV ratios.  These results demonstrate that a variable IMF alone has difficulty explaining the observed scatter in the \halpha-to-FUV ratios.  We conclude by considering the limitations of the model SFHs, and discuss the use of additional empirical constraints to improve future SFH modeling efforts.

\end{abstract}


\section{Introduction}

Two of the most widely used tracers of recent star formation are the nebular \halpha\ recombination line and the ultra-violet (UV) continuum.  The \halpha\ arises from recombination of gas ionized by photons from massive stars ($\gtrsim$ 15 \msun) and is expected to be observed over the typical lifetimes of extremely massive stars ($\lesssim$ 5 Myr).  The UV continuum is due to non-ionizing photospheric emission from  stars with $M$ $\gtrsim$ 3 \msun, which have lifetimes  $\lesssim$ 300 Myr.  In tandem, these two integrated tracers provide leverage on current and recent star formation in both nearby and distant galaxies (see \citealt{ken98} and references therein).

In principle, observed \halpha\ and UV luminosities should yield consistent measures of star formation. Under the assumption of a constant SFH over a sufficiently long timeline (e.g., $\sim$ 1 Gyr), the expected ratio of \halpha\ and UV star formation rates (SFRs) should be constant with respect to both time and environment (see \citealt{ken98} and references therein), assuming a fully populated and universal IMF, solar metallicity, all Lyman continuum photons ionize hydrogen, and with no attenuation due to dust. Deviations from this fiducial ratio would suggest that one or more of the underlying assumptions are not correct. 

A number of studies have demonstrated discrepancies in SFRs as measured by \halpha\ and UV luminosities \citep[e.g.,][]{bua87, bua92, gla99, yan99, sul00, bel01b, moo00, sul04, igl04}.  Recently, several studies have unambiguously shown that the observed discrepancy in \halpha\ and UV SFR indicators is systematic, such that the observed ratio of \halpha-to-UV flux declines with decreasing galaxy luminosity \citep[e.g.,][]{meu09, lee09b, bos09}.  

\begin{deluxetable*}{lcccccc}
\tablecolumns{7}
\tablecaption{Basic Properties of the LVL Galaxies}

\tablehead{   
\colhead{Galaxy} &   
    \colhead{$\log(M_{\star})$} &
    \colhead{$\log(\Sigma_{R})$}  &
    \colhead{$M_{B}$}  &
    \colhead{$\log[SFR(H\alpha)]$}  &
    \colhead{$\log \frac{F(H\alpha)}{F(FUV)} + \kappa$}  &
    \colhead{Source}  \\
    \colhead{Name} &   
    \colhead{$\log(M_{\odot})$} &
    \colhead{$\log(L_{\odot}~kpc^{-2})$}  &
    \colhead{}  &
    \colhead{(\msun~yr$^{-1}$)}  &
    \colhead{}  &
    \colhead{}  \\
     \colhead{(1)} &   
    \colhead{(2)} &
    \colhead{(3)}  &
    \colhead{(4)}  &
    \colhead{(5)}  &
    \colhead{(6)}  &
    \colhead{(7)}  \\
       }
      
\startdata    
UGCA292 &  5.65 &  6.91 &   -11.42 & -2.755 &  0.031 &        1 \\
UGC5364 &  5.82 &  7.24 &   -11.40 & -4.135 & -0.739 &        1 \\
UGC8091 &  5.84 &  7.54 &   -12.07 & -2.700 & -0.007 &        1 \\
UGCA438 &  6.28 &  7.65 &   -12.35 & -4.316 & -1.668 &        2 \\
UGC8833 &  6.44 &  7.60 &   -12.42 & -3.331 & -0.468 &        1 \\
UGC4483 &  6.50 &  7.72 &   -12.86 & -2.446 &  0.016 &        2 \\
UGC9128 &  6.51 &  7.56 &   -12.39 & -3.970 & -0.948 &        1 \\
CGCG269-049 &  6.52 &  7.88 &   -12.35 & -3.120 & -0.290 &        1 \\
KKH37 &  6.53 &  7.56 &   -11.67 & -3.818 & -0.319 &        2 \\
UGCA281 &  6.54 &  8.32 &   -13.50 & -1.391 &  0.369 &        1

\enddata
\tablecomments{The observed and derived properties of the 185 galaxies considered in this study.  For brevity, we list 10 here and make the rest available in machine readable format.  The measurement of galaxy stellar masses and $R$-band surface brightnesses are detailed in \S \ref{data}.  The \halpha\ SFR (column (4) have not been corrected for attenuation, while the \halpha-to-FUV ratios in column (6) have been dust corrected as detailed in \S \ref{dustcor}.  The values in this column are listed relative to the value of fiducial, whose value is reflected by the constant $\kappa =$ $-$13.17. Column (7) indicates the source of the optical observations, either from SDSS DR7 \citep{aba09} (1) or LVL \citep{vanzeeprep} (2).  The full data table will be available with the journal version of the paper or upon request.}
\label{tab1}
\end{deluxetable*}

Despite extensive research, there remains no consensus for the cause of this trend.  Independent studies have verified that factors such as assumed metallicity and choice of stellar models cannot be responsible for the observed trend \citep[e.g.,][]{meu09, lee09b, bos09}, while investigations of other contributing factors, including non-constant SFHs, a variable IMF, and ionizing photon leakage have yet to yield conclusive results.

In this paper, we undertake a focused investigation on the impact of non-constant SFHs on observed \halpha-to-UV ratios.  Our main objective is to find a set of simple SFH models (i.e., periodic bursts of star formation superimposed on a baseline constant SFR) that are well matched to the observed distribution of \halpha-to-UV SFR ratios, while being consistent with other available data.  To do this, we compare the observed distribution of \halpha-to-FUV ratios to the predicted distributions from a set of simple SFH models, in bins of galaxy stellar mass.  Using the results of this comparison, we examine the ability of the highest probability model SFHs to explain the observed trend in \halpha-to-FUV flux ratios versus stellar mass, $R$-band surface brightness, \halpha\ luminosity, and $M_{B}$.   In addition, we compare SFH model parameters and predictions with previous studies that have analyzed trends in \halpha-to-UV ratios using different SFHs or IMF assumptions.  We conclude the paper by discussing the strengths and limitations of simple model SFHs and suggest specific ways to improve future SFH modeling efforts.

\begin{figure}[b]
\begin{center}
\epsscale{1.2}
\plotone{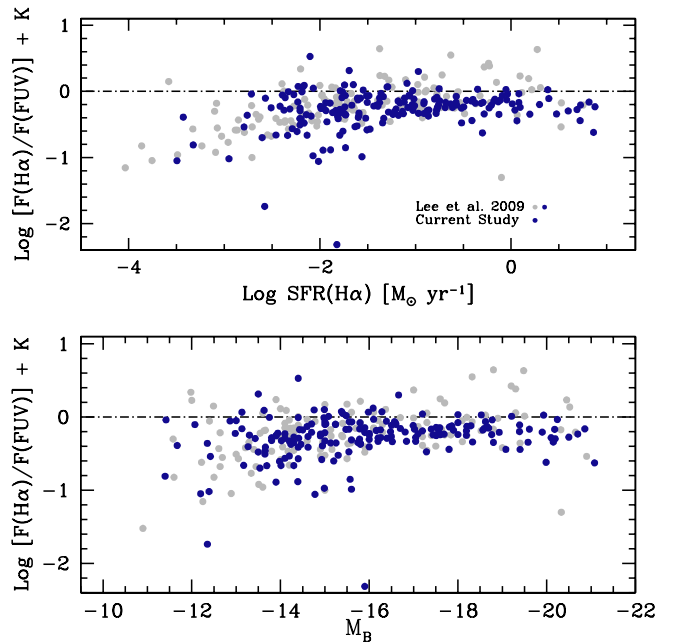}
\caption{\scriptsize \halpha-to-FUV ratios plotted versus the \halpha\ SFR and $M_{B}$ for the \citet{lee09b} sample (grey and navy points) and the sample considered in this study (navy points only).  All ratios have been corrected for both foreground and internal extinction using the method of \citet{lee09b}.  To help illustrate the differences in the flux ratios, a constant $\kappa$ has been added.  The value of $\kappa$ is $-$13.17, the negative of the expected flux ratios from our models for a constant SFH.  The black dot-dashed line indicates the expected fiducial value for a constant SFH, solar metallicity, and fully populated Chabrier IMF. The two samples are generally consistent, although there is a noticeable deficiency in luminous galaxies ($M_{B}$ $\lesssim$ $-$16) with \halpha-to-FUV ratios above the fiducial in the present study, relative to the larger sample of \citet{lee09b}.} 
\label{janicesample}
\end{center}
\end{figure}

\section{The Observational Data}
\label{data}

\begin{deluxetable*}{lccc|cc|ccc}
\tablecolumns{9}
\tabletypesize{\scriptsize}
\tablewidth{0pt}

\tablecaption{\halpha-FUV Flux Ratio Distribution Statistics for Stellar Mass Bins}
\tablehead{   
\colhead{$\log SFR(H\alpha)$} &   
\colhead{$\log \frac{SFR(H\alpha)}{SFR(FUV)}$} &
    \colhead{1$\sigma$-scatter} &
    \colhead{N$_{gal}$} &
\colhead{$\log \frac{SFR(H\alpha)}{SFR(FUV)}$} &
    \colhead{1$\sigma$-scatter} &
    \colhead{$\log \frac{SFR(H\alpha)}{SFR(FUV)}$} &
    \colhead{1$\sigma$-scatter} &
    \colhead{N$_{gal}$} \\
     \colhead{(1)} &   
\colhead{(2)} &
    \colhead{(3)} &
    \colhead{(4)} &   
\colhead{(5)} &
    \colhead{(6)} &
    \colhead{(7)} &   
\colhead{(8)} &
    \colhead{(9)} 
    }
    
    \startdata
    0.50 & -0.14 & 0.29 & 20 & -0.32 & 0.16 & -0.06 & 0.05 & 7\\
    -0.25 & -0.09 & 0.20 & 26 & -0.28 & 0.13 & -0.03 & 0.08 & 22\\
    -0.75 & -0.13 & 0.22 & 44 & -0.22 & 0.15 & -0.06 & 0.10 & 21\\
    -1.25 & -0.13 & 0.22 & 51 & -0.19 & 0.15 & -0.06 & 0.14 & 36\\
    -1.75 & -0.18 & 0.17 & 58 & -0.24 & 0.26 & -0.14 & 0.24 & 38\\
    -2.25 & -0.27 & 0.22 & 56 & -0.38 & 0.45 & -0.26 & 0.44 & 40\\
    -2.75 & -0.51 & 0.25 & 31 & -0.38 & 0.45 & -0.26 & 0.44 & 15\\
    -3.50 & -0.55 & 0.57 & 21 & -0.69 & 0.31 & -0.53 & 0.26 & 6
\enddata

\tablecomments{A comparison between the \halpha-to-FUV ratios as a function of \halpha\ SFR between the current sample and that of \citet{lee09b}.  Columns (1)-(4) are the dust corrected \halpha\ SFRs, \halpha-to-FUV ratios, 1$\sigma$-scatter in the ratio and number of galaxies per bin taken from Table 2 in \citet{lee09b}.  Columns (5) and (6) are the \halpha-to-FUV ratio and 1$\sigma$ scatter for galaxies in the present study, with the extinction corrections of \citet{lee09b} applied.  Columns (7) and (8) are the \halpha-to-FUV ratio and 1$\sigma$ scatter for galaxies in the present study, with the extinction corrections from \citet{ken09} and \citet{hao11} applied.  Column (9) is the number of galaxies per bin for the sample of 185 galaxies we consider in this study.  }
\label{samplediffs}
\end{deluxetable*}

We test our SFH models on a sample of 185 nearby star-forming galaxies, whose observable properties are listed in Table \ref{tab1}.  To arrive at this particular sample, we began by considering the 390 galaxies with \halpha\ and GALEX FUV observations from the 11 Mpc \halpha\ and UV Galaxy Survey \citep[11HUGS;][]{ken08, lee11}.  We further restricted the sample to those galaxies that also had high quality ancillary data to To allow us  correct for the effects of dust and study the \halpha-to-FUV ratio trends versus galaxy stellar mass and surface brightness.  Specifically, we selected 11HUGS galaxies that are also members of the smaller \emph{Spitzer} Local Volume Legacy survey \citep[LVL; e.g.,][]{dale09}.  The LVL sample contains 258 galaxies with comprehensive \emph{Spitzer} IR imaging.  However, only 59 galaxies in the sample also have homogeneous optical broadband imaging \citep{vanzeeprep}.  To increase the sample size, we also considered LVL galaxies that fall within the footprint of SDSS DR7 \citep{aba09}.  There are 126 additional galaxies that met this requirement.  Thus, we have a final sample of 185 galaxies with observations of \halpha\ and GALEX FUV from 11HUGS, \emph{Spitzer} IR from LVL, and  optical broadband imaging from either SDSS or LVL. For consistency with the LVL optical imaging, fluxes measured from SDSS images were converted to the Johnson filter system using the transformations provided in \citet{bla05}.

In Figure \ref{janicesample}, we show a comparison between the \halpha-to-FUV ratios from \citet{lee09b} (navy and grey points) with those of the 185 galaxies considered in this study (navy points only). The \halpha, FUV, and B-band fluxes, as well as the adopted dust corrections in this figure have all been taken from \citet{lee09b}.  

Although similar in most aspects, there are two notable difference between the samples.  First, our subset of 185 galaxies are noticeably deficient in luminous systems ($M_{B}$ $\sim$ $-$16) with \halpha-to-FUV ratios significantly greater than the fiducial.  Second, our sample contains fewer low luminosity systems than the larger sample of \citet{lee09b}.  In both cases, the limiting factor is the lack of sufficient ancillary data.  In the case of the low luminosity galaxies, nearly all that have been excluded from the present study lack sufficiently homogenous optical broadband coverage.  For more luminous galaxies, some lack either ancillary optical imaging or fall outside the boundaries of the LVL survey and therefore lack \emph{Spitzer} imaging.

We next examine the differences in \halpha-to-FUV ratios between the two samples.  In Table \ref{samplediffs}, we quantify the \halpha-to-FUV ratios as a function of \halpha\ SFR, after applying identical extinction corrections from \citet{lee09b} (see \S \ref{duscor}).  As expected, the largest difference is seen at high \halpha\ SFR, where our sample lacks galaxies with high \halpha-to-FUV ratios.  For $\log[SFR(H\alpha)]$ $\gtrsim$ $-$1.75, the mean \halpha-to-FUV ratios in the present study are systematically $\sim$ 0.2 dex lower than those in the sample of \citet{lee09b}.  Similarly, the 1$\sigma$ scatter over the same range is $\sim$ 0.1 dex lower in the present sample.  However, for galaxies with $\log[SFR(H\alpha)]$ $\lesssim$ $-$1.75, the \halpha-to-FUV ratios have comparable mean values and scatter, per bin of \halpha\ SFR.  Outside of the differences between the samples  at high \halpha\ SFRs, the two samples are in general agreement.  This indicates that possible selection effects associated with the smaller sample in this study are minimal.

\subsection{Adopted Extinction Corrections}
\label{dustcor}

For accurate comparison of \halpha\ and FUV SFR indicators, corrections are required for extinction due to both Galactic foreground and to internal effects in each galaxy.  We correct for foreground extinction using the maps of \citet{sch98}, assuming the Milky Way extinction law of \citet{car89}.  For internal attenuation, we use two types of corrections: the first based on direct IR measurement of reprocessed light and another using correlation-based prescriptions (e.g., see Appendix D in \citealt{ler08}).

For $\sim$ 70\% of the sample, we have estimates of the total-IR flux based on Spitzer/MIPS far-IR (FIR) measurements, allowing us to use the `energy balance' extinction correction methods derived by \citet{ken09} for \halpha\ and by \cite{hao11} for FUV.  These methods derive an intrinsic luminosity from the combination of unobscured (e.g., \halpha, FUV) and obscured (i.e., FIR continuum) signatures of star formation.  From the intrinsic luminosity one can correct for the effects of internal dust attenuation on the observed \halpha\ and FUV fluxes.  For the $\sim$ 30\% of the sample where FIR measurements were not available, we used the same correlation-based attenuation corrections as \citet{lee09b}, namely a $B$-band luminosity prescription for $A_{H\alpha}$ and a dust-law prescription for $A_{FUV}$.  While both methods are anchored to Balmer decrement extinction corrections, the self-consistency of the `energy balance' method along with the availability of IR for the majority of our sample make it an appealing choice for this study.  

More explicitly, the `energy balance' extinction correction for $A_{H\alpha}$ from \citet{ken09} is:

\begin{equation}
A_{H\alpha} = 2.5 \log \left( 1 + \frac{0.0019 \ L( \mbox{TIR})}{L(H\alpha)_{obs}} \right)
\end{equation}
 
\noindent and from \citet{hao11}, the correction for the FUV is given as: 
 
\begin{equation}
A_{FUV} = 2.5 \log \left( 1 + \frac{0.37 \ L( \mbox{TIR})}{L(FUV)_{obs}} \right)
\end{equation}

\noindent where the TIR luminosity is a MIPS-based measurement made using the conversion in \citet{dal02}:

\begin{equation}
\begin{split}
L( \mbox{TIR}) = 1.559 \nu f_{\nu}(24 \mu m) + 0.7686 \nu f_{\nu}(70 \mu m) \\
+ 1.347 \nu f_{\nu}(160 \mu m)
\end{split}
\end{equation}

\begin{figure}[t]
\begin{center}
\epsscale{1.2}
\plotone{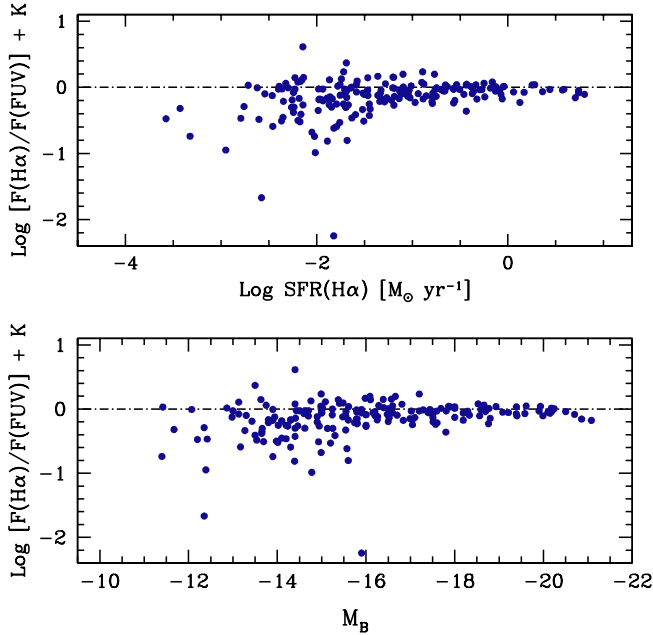}
\caption{Same as Figure \ref{janicesample}, except that we have applied the dust correction of \citet{ken09} for \halpha\ and \citet{hao11} for the FUV.  We adopt these dust corrections for all subsequent analysis in this paper.}
\label{hasfrmb}
\end{center}
\end{figure}

Overall, our internal extinction corrections yield similar results to those used by \citet{lee09b}.  As shown in Table \ref{samplediffs}, the main difference between the two sets of corrections is seen for the most luminous galaxies in the sample.  Here, \citet{lee09b} find the mean \halpha-to-FUV ratio to be $\sim$ 0.2-0.3 dex below the expected fiducial, whereas the energy balance dust corrections place the mean ratios within $\sim$ 0.05 dex of the fiducial.  However, for galaxies with low \halpha\ SFRs, the adopted extinction correction does not drastically influence the mean or scatter in the \halpha-to-FUV ratios; these galaxies have low dust contents, such that extinction corrections have only modest impacts on their luminosities.  We plot the flux ratios for our sample with the energy balance extinction corrections applied in Figure \ref{hasfrmb}.  

Although differences in dust corrections are both important and interesting in their own right, analyzing the impact of various extinction corrections on SFR indicators is beyond the scope of the present analysis.  Instead, we refer the reader to \citet{lcj4prep}, which investigates the reliability and consistency of various \halpha\ and FUV extinction corrections, including those from \citet{lee09b}, \citet{ken09}, and \citet{hao11}.

\subsection{Deriving Galaxy Masses}

The stellar mass for each galaxy in the sample was derived by spectral energy distribution (SED) fitting as detailed in \citet{johnprep}, which closely follows the methods detailed in \citet{sal07}.  Briefly, we fit the observed  UV, optical data, and IR luminosities of each galaxy with a suite of  \citet{bru03} population synthesis models, whose parameters span a range in stellar metallicity, age, exponentially declining SFHs, and dust properties.  For each model, we derive a stellar mass from the best fitting normalization of the model SED. We also construct the cumulative distribution function for the stellar mass  from the likelihoods of the normalized model SEDs for each sample galaxy.  The stellar masses reported here are the medians of the cumulative distribution function.  In practice, the stellar mass is surprisingly robust,  as the effects of dust and age on the optical mass-to-light ratio have correspondingly similar effects on the observed optical color \citep{bel01}, while the effects of stellar metallicity on the optical mass-to-light ratios are minimal.  The derived stellar masses for each system are presented in Table \ref{tab1}.

\begin{figure}[t]
\begin{center}
\epsscale{1.2}
\plotone{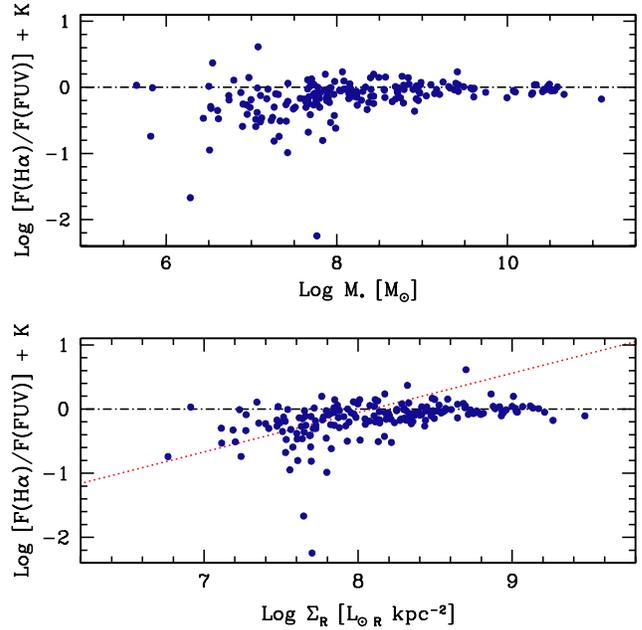}
\caption{Same as Figure \ref{janicesample} only plotted versus galaxy stellar mass and $R-$band surface brightnesses ($\Sigma_{R}$). In the lower panel, we have over-plotted the line fit line to the data studied by \citet{meu09} as the red dotted line. The \halpha-to-FUV ratios considered in this study do not show as strong a correlation with $\Sigma_{R}$.}
\label{obs_data}
\end{center}
\end{figure}

The \halpha-to-FUV ratio as a function of galaxy stellar mass (Figure \ref{obs_data}) shows subtle, but important differences compared to when the ratios are plotted versus \halpha\ SFR or $M_{B}$, as in Figure \ref{hasfrmb}. For example, when considering the ratios versus \halpha\ SFR, we see there are no galaxies with \halpha-to-FUV ratios above the fiducial for $\log[H\alpha(SFR)]$ $\lesssim$ $-$2.5.  Additionally, there are only a few galaxies with \halpha-to-FUV ratios near or above the fiducial for systems with $M_{B}$ $\lesssim$ $-$13.  From these two panels, one could conclude that galaxies with low luminosities or low current SFRs must have systematically low \halpha-to-FUV ratios.

However, when plotting the same ratios versus stellar mass, one could reach a different conclusion.  As shown in the top panel of Figure \ref{obs_data}, a number galaxies with masses $\lesssim$ 10$^{7}$ \msun\ have \halpha-to-FUV ratios near or above the fiducial.  Recent bursts of SF have little impact on a galaxy's total mass, but can strongly influence the observed \halpha\ or optical luminosities \citep[e.g.,][]{bel01}.  Thus, considering the data versus total stellar mass minimizes the covariances that affect the \halpha-to-FUV ratios when plotted versus luminosity.

\subsection{Measuring $R$-band Surface Brightness}

A similar study conducted by \citet{meu09} found a strong trend between $R$-band surface brightness ($\Sigma_{R}$) and \halpha-to-FUV ratios.  To facilitate a direct comparison, we utilize $R$-band imaging from either SDSS or LVL, and follow the surface brightness measurement procedure described in \citet{meu06}.  Below, we briefly summarize the methodology.

We first masked foreground stars and background galaxies identically to the method described in \citet{dale09}.  With these objects removed, we computed the surface brightness profiles using elliptical apertures.  The major and minor axes of a measurement aperture are scaled, preserving ellipticity, such that the ellipse encloses 50\% of the total flux.  We then normalized the total flux to the area of the aperture to derive the surface brightness.  Particular care was taken to ensure that these scaled measurement apertures, constructed using ellipticity and position data compiled from NED/SIMBAD, were properly shaped and centered. For consistency with \citet{meu09}, we note that the flux percentage choice of 50\% rather than 90\% does not affect the overall trends in our data, beyond shifting to slightly higher surface brightnesses.

We plot the \halpha-to-FUV ratios versus $R$-band surface brightness ($\Sigma_{R}$) in the bottom panel of Figure \ref{obs_data}.   Again, we see a decline in \halpha-to-FUV ratios toward lower surface brightness systems.  In this case, the systematic decline of \halpha-to-FUV ratios is not as strong as when the sample is plotted versus \halpha\ SFR, but is still more evident than when the ratios are considered versus galaxy stellar mass. Like all optical luminosities, $R$-band fluxes are known to vary due to episodes of recent SF \citep[e.g.,][]{bel01}, which likely explains at least some of the correlation between \halpha-to-FUV ratios and $\Sigma_{R}$.  We note that the observed trend between \halpha-to-FUV ratio and $R$-band surface brightness in our sample is not quite as strong as that presented in \citet{meu09}, which we have over-plotted as the red dotted line in Figure \ref{obs_data}.  We discuss some of the potential reasons for this disagreement in \S \ref{previous}. 

\begin{deluxetable}{lcc}
\tablecolumns{3}
\tablewidth{0pt}

\tablehead{   
\colhead{Bin} &   
\colhead{$\log \frac{SFR(H\alpha)}{SFR(FUV)}$} &
    \colhead{N$_{gal}$} \\
    \colhead{$\log(M_{\odot})$} &   
\colhead{Median} &
    \colhead{} \\
    \colhead{(1)} &   
\colhead{(2)} &
    \colhead{(3)} \\
      }
      \tablecaption{\halpha-FUV Flux Ratio Distribution Statistics for Stellar Mass Bins}
\startdata    
 $\log(M_{\star})$ $>$10.0  & -0.05$_{-0.06}^{0.00}$  & 16\\
 10.0$\ge$$\log(M_{\star})$$>$9.0 & -0.01$_{-0.09}^{+0.07}$ & 20\\
9.0$\ge$$\log(M_{\star})$$>$8.0 &-0.06$_{-0.14}^{+0.14}$ & 55\\
 8.0$\ge$$\log(M_{\star})$$>$7.0 &-0.17$_{-0.34}^{+0.17}$ & 71 \\
 $\log(M_{\star})$ $\le$7.0 &-0.27$_{-0.32}^{+0.36}$ & 23
 \enddata
\tablecomments{The median \halpha-to-FUV ratio per stellar mass bin.  The errors bars correspond to the 16th and 84th percentile values of the distribution of observed ratios.  The number of galaxies in each bin is listed in column (3).}
\label{tab2}
\end{deluxetable}

\section{Model Star Formation Histories}

The model SFHs we consider in this analysis are simple toy models that are primarily composed of periodic bursts superimposed onto a constant SFR.  Given that modest variations in the period, duration, and amplitude of bursts can result in significantly different \halpha, UV, and broadband fluxes, we consider a diverse and extensive set of $\sim$ 1500 model SFHs, whose properties and construction we describe below.

\subsection{Construction of the Model Star Formation Histories and Associated Fluxes}
\label{mod_sfhs}

We model SFHs that focus on two separate regimes: the ancient epoch ($>$ 1 Gyr ago) which has little impact on the \halpha-to-FUV ratios and recent SFHs ($<$ 1 Gyr ago) which are characterized by a mix of constant SF and bursts.  We have modeled the older stellar populations that dominate a galaxy's total stellar mass using a 1 Gyr burst of SF 8 Gyr ago to populate the red giant branch and ancient main sequence, as well 1 Gyr of constant SFR from 1-2 Gyr ago to account for the contribution of asymptotic giant branch stars.  The precise epoch of the ancient and intermediate age SFHs do not significantly change any of our analysis.  Changing the ages of these older bursts results in small variations in the absolute values of $\log(\Sigma_{R})$, i.e., $\lesssim$ 0.5 dex.  However, because the same ancient and intermediate SFHs are used for all models, there is no effect on the relative values of $\log(\Sigma_{R})$.  Finally, to match findings from measured SFHs in nearby dwarf galaxies \citep[e.g.,][]{dol05, wei11}, we scaled the SFHs such that SF from $>$ 1 Gyr ago accounts for 90\% of the total stellar mass formed in each model.

\begin{figure}[b]
\begin{center}
\epsscale{1.2}
\plotone{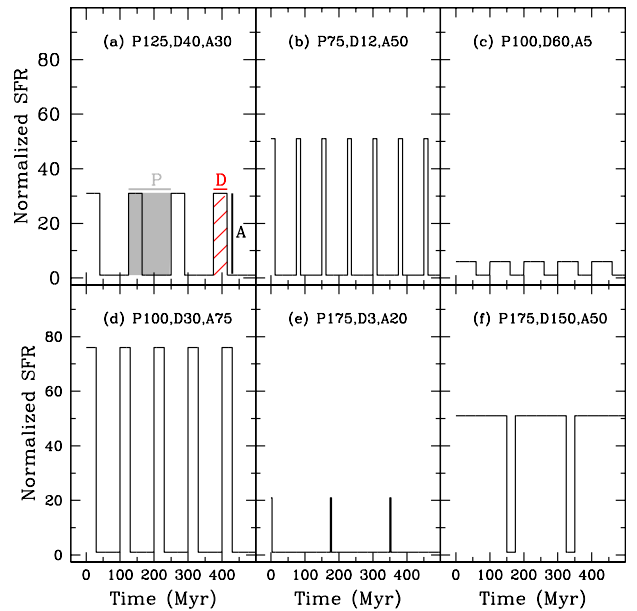}
\caption{An illustrative sampling of select model SFHs we consider in this paper, plotted over a 500 Myr interval. In panel (a) we have denoted the period (P) of a SFH cycle by the grey shaded region, the duration (D) of a burst by the red-hatched region, and the amplitude (A) is the increase in SFR above the constant equilibrium value. Throughout the paper we refer to the models with nomenclature such as P125, D40, A30, which corresponds to a period of 125 Myr, a duration of 40 Myr, and amplitude of 30, the model SFH in panel (a).  Panel (f) is an example of a `gasping' SFH.  In total we have constructed $\sim$ 1500 model SFHs as described in \S \ref{mod_sfhs}. }
\label{sample_sfhs}
\end{center}
\end{figure}

We constructed the recent model by assuming a constant SFR over the past 1 Gyr and superimposing bursts of SF.  We assume the bursts to be a 1 Myr long episode of constant SF, which we then interpolated onto a finer time grid to track the evolution of \halpha\ and UV fluxes over longer timescales.  This basis model was then shifted in time, multiplied by the appropriate amplitude, and finally linearly co-added with other basis models to obtain the desired recent SFH.

Each model has a specified burst amplitude (A; the ratio of maximum SFR to constant SFR), burst duration (D), and period between bursts (P). We additionally characterize our models by the quantity ($D$$\times$$A$)/$P$, which is the ratio of stellar mass formed during a single burst to the mass formed assuming the baseline SFR over one period.  Selected permutations of A (2, 5, 10, 15, 20, 30, 40, 50, 75, 100, 150), D (3, 6, 12, 18, 25, 30, 40, 50, 60, 75, 100, 150, 200 Myr), and P (5, 10, 15, 20, 25, 30, 40, 50, 60, 75, 100, 125, 150, 175, 200, 225, 250 Myr) combined with the constraint $P>D$ resulted in 1441 different model SFHs.  We added 25 more models in which we allowed the SFR between bursts to go to zero, i.e., the burst amplitude is infinite.  We also consider a constant SFH model (i.e., A $=$ 1). Many of the models we consider fall under the literature parlance of `bursting' or `gasping' SFHs \citep[e.g.,][]{ann03, meu09}.  Although the definition of both terms is not universal, here we consider a `bursting' model to have $A>1$ and $P>>D$, while for a `gasping' model $D$ $\sim$ $P$ with the deficit in SF equal to 1/A.  That is, a gasp is essentially a decrease in the SFR from an otherwise constant rate.  We show a select sample of model SFHs in Figure \ref{sample_sfhs}.

\begin{figure}[t]
\includegraphics[scale=0.35, angle=270]{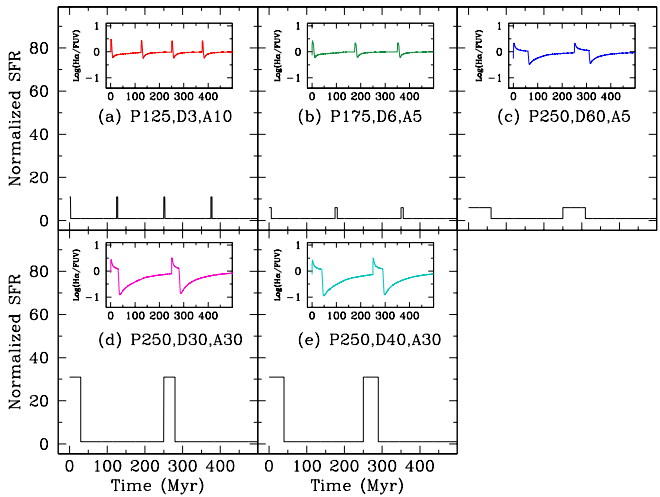}
\caption{The model SFHs that best describe the \halpha-to-FUV flux ratio distribution in order of decreasing bins of stellar mass ((a) $\log(M_{\star})$ $>$10.0, (b) 10.0$\ge$$\log(M_{\star})$$>$9.0, (c) 9.0$\ge$$\log(M_{\star})$$>$8.0, (d) 8.0$\ge$$\log(M_{\star})$$>$7.0, (e) $\log(M_{\star})$ $\le$7.0; see Table \ref{tab3}), as determined by a two sided KS test.  The inset shows the evolution of the \halpha-to-FUV SFR ratio as a function of time for each model SFH.  The red, green, blue, magenta, and cyan color coding each refer to the best fit SFH models in the bins of highest to lowest stellar mass, respectively.  This color coding scheme will be used throughout the remainder of the paper in reference to a particular SFH model.}

\label{best_fit_sfhs}
\end{figure}

Synthesizing the evolution of \halpha, FUV, and $R$-band fluxes was done consistently with the construction of the model SFHs. A 1 Myr long burst of SF was input into the synthesis code of \citet[][]{bru07}.  The resulting spectra were convolved with the appropriate filter response functions,  and the resulting fluxes were linearly co-added identically to the SFHs.  In calculating these integrated quantities, we selected the Padova stellar evolution models \citep{bre93, fag94}, solar metallicity, and a fully populated Chabrier IMF with mass limits of 0.1 and 100 \msun.  For consistency with the observations, we convolved the UV spectrum with the GALEX FUV filter response function.  The \halpha\ flux is modeled as the number of ionizing photos divided by $10^{11.87}$, the expectation value for Case B recombination.  We do not include losses due to the leakage or absorption of ionizing photons by dust within the Str\"{o}mgren sphere. In these models, the \halpha\ and FUV fluxes provide for consistent measures of SF when averaged over the course of 1 Gyr, i.e., the ratio of the sums, $\sum$(\halpha)/$\sum$(UV), is constant, even though many individual points may deviate from this average.

Throughout the paper, we assume a fiducial value of \halpha-to-FUV ratios based on the scenario in which a galaxy which has been forming stars at a constant rate over its entire lifetime.  For clarify in plotting, we have set the logarithm of the fiducial \halpha-to-FUV ratio equal to zero.  Consequently, all model and observed \halpha-to-FUV ratios have been adjusted appropriately such that $\log[F(H_{\alpha})/F(FUV)]_{plotted}$ $=$ $\log[F(H_{\alpha})/F(FUV)] + \kappa$, where $\kappa =$ $-$13.17

To derive the model $R$-band surface brightness values, we converted the $R$-band fluxes to a surface brightness assuming an area of 1 kpc$^{2}$.  In practice, this areal normalization will increase for more massive galaxies.  However, as we are only interested in variations in the logarithmic surface brightnesses, altering the areal normalization by a factor of a few has minimal impact on the placement of the models. Overall, this method results in model $R$-band fluxes (and surface brightnesses) that are consistent with observations of galaxies with these typical SFRs.

Uncertainties in the observed \halpha, FUV, and $R$-band fluxes are not inherently factored into the modeled fluxes.  As this may contribute to the observed scatter, we applied independent small random shifts based on observational uncertainties in the \halpha\ and FUV fluxes listed in \citet{ken08} and \citet{lee11}, to each of the model flux values.  Specifically, we drew a set of errors from a randomly generated normal distribution with a mean of 0 and a dispersion of 0.05 mag.    

\begin{deluxetable*}{lcccccc}
\tablecolumns{7}
\tablewidth{0pt}

 \tablecaption{Model KS Test Probabilities}

\tablehead{   
\colhead{Bin} &   
\colhead{Model} &
    \colhead{Period} &
    \colhead{Duration} &
       \colhead{Amplitude} &
       \colhead{$\frac{D\times A}{P}$} &
       \colhead{KS Probability}\\
       \colhead{$\log(M_{\odot})$} &   
\colhead{} &
    \colhead{(Myr)} &
    \colhead{(Myr)} &
       \colhead{} &
       \colhead{} &
       \colhead{(\%)} \\
       \colhead{(1)} &   
\colhead{(2)} &
    \colhead{(3)} &
    \colhead{(4)} &
       \colhead{(5)} &
       \colhead{(6)} &
       \colhead{(7)}
       }
      
\startdata    
$\log(M_{\star})$ $>$10.0 & Best & 125 & 3 & 10 & 0.24 &  72.3 \\
& 10th Best & 75 & 12 & 2 & 0.32 & 65.6 \\
& 10th Worst & 125 & 12 & 150 & 14.4 & 4.9$\times$10$^{-10}$ \\
10.0$\ge$$\log(M_{\star})$$>$9.0 & Best & 175 & 6 & 5 & 0.17 & 99.9 \\
& 10th Best & 225 & 6 & 5 & 0.13 & 99.6 \\
& 10th Worst & 75 & 3 & 100 & 4.00 & 2.53$\times$10$^{-12}$ \\
9.0$\ge$$\log(M_{\star})$$>$8.0 & Best & 250 & 60 & 5 & 1.20 & 94.3 \\
& 10th Best & 225 & 40 & 5 & 0.89 & 83.4 \\
& 10th Worst & 75 & 3 & 150 & 6.00 & 1.67$\times$10$^{-31}$ \\
8.0$\ge$$\log(M_{\star})$$>$7.0 & Best & 250 & 30 & 30 & 3.60 & 97.5 \\
& 10th Best & 250 & 25 & 40 & 4.00 & 73.4 \\
& 10th Worst & 75 & 12 & 100 & 16.0 & 1.1$\times$10$^{-22}$\\
$\log(M_{\star})$ $\le$7.0 & Best & 200 & 40 & 30 & 6.00 & 99.8 \\
& 10th Best & 150 & 30 & 20 & 4.00 & 95.8 \\
& 10th Worst & 250 & 6 & 2 & 0.05 & 3.6$\times$10$^{-7}$

\enddata
\tablecomments{Parameters and KS probabilities for the best, 10th best, and 10th worst fit model SFHs in each bin of stellar mass.}
\label{tab3}
\end{deluxetable*}

\section{Results}

\subsection{Comparing Modeled and Observed \halpha-to-FUV Ratios}
\label{results}

The goal of our analysis is to identify a set of characteristic SFHs that can match the observed distribution of \halpha-to-FUV ratios within bins of  galaxy stellar masses, without violating other observational constraints. To determine this characteristic SFH, we compare the observed distribution of \halpha-to-FUV ratios with predicted distributions from each of the model SFHs.

We first divide the observational sample into bins of stellar mass, under the assumption that galaxies with comparable stellar masses are likely to share a characteristic SFH. We chose a simple scheme that consists of five bins with sizes of $\Delta$$\log(M_{*}) \sim$ 1.  The resulting division provides a reasonable balance between grouping galaxies of similar masses and populating each bin with an adequate number of data points.  The number of galaxies, median \halpha-to-FUV ratio, and scatter in the ratio for each bin are listed in Table \ref{tab2}.

As a metric for comparison between the models and observed \halpha-to-FUV ratios, we have adopted a two-sided Kolmogorov-Smirnov (KS) test.  Unlike other statistical tests (e.g., $t$-test, Anderson-Darling), the KS test makes no assumption about the form of the parent distribution of either population, i.e., they need not be normally distributed.  Indeed, the distributions of the \halpha-to-FUV ratios are generally complex and difficult to parameterize, making the KS test better suited for this particular analysis.  For simplicity, we will adopt multiples of the typical $\sigma$ values to evaluate the quality of the model-data agreement (e.g., 1-$\sigma$ $\sim$ 68\%, 2-$\sigma$ $\sim$ 95\%, etc).  We also include parameters and KS test probabilities for the 10th best and 10th worst fits to highlight the differences in the model parameters in the high and low probability regimes.

The results of the model-data comparison are summarized in Table \ref{tab3}.  In each bin of stellar mass, we have indicated the highest probability model parameters (i.e., $P$, $D$, $A$), the value of ($D$$\times$$A$)/$P$, and the KS probability. In general, in each bin of stellar mass, our highest probability models provide reasonable matching to the observed distributions of \halpha-to-FUV ratios, as judged by the KS test.  Four of the five bins have models that match the data with a probability of $\gtrsim$ 94\%.  The lowest mass bin ($\log(M_{\star})$ $\le$7.0) and second highest mass bin  (10.0$\ge$$\log(M_{\star})$$>$9.0) have KS probabilities of 99.8\% and 99.9\%, respectively.  The lowest probability model is in the highest mass bin ($\log(M_{\star})$ $>$10.0), which has a KS probability of 72.3\%.  Despite the simplicity of the SFH models considered, we find all highest probability models agree with the data at levels of $\gtrsim$ 1-3 $\sigma$.

Examining the highest probability model parameters in each stellar mass bin, we see the SFHs of higher mass systems are generally characterized by constant SFHs with an occasional modest burst.  Specifically, the two highest mass bins favor models  of short duration ($D$ $\lesssim$ 6 Myr) and relatively modest amplitudes ($A$ $\lesssim$ 10).  These SFHs are predominantly constant SFRs, with bursts interspersed $\sim$ 5\% of the time. The central bin (9.0$\ge$$\log(M_{\star})$$>$8.0) is best characterized by a long period ($P =$ 250 Myr) of constant SF and an increase in amplitude by a factor of 5 for a 60 Myr duration.  The two lowest mass bins have highest probability models with higher amplitude bursts ($A$ $\sim$ 30), relatively long durations ($D$ $\sim$ 30-40 Myr), and large periods ($P =$ 250 Myr). The highest probability SFH models along with the corresponding temporal evolution of the \halpha-to-FUV ratios are shown in Figure \ref{best_fit_sfhs}.

As a rough proxy for the robustness of the highest probability model, we rank the models according to the probability of matching the data, as judged by the KS test.  We then examine 10th most probable models in each stellar mass bin and find they have similar parameters to the corresponding highest probability models.  The largest contrast between the best and 10th best models is in the highest mass bin, where the duration increases by a factor of 4 (from 3 to 12) and the amplitude decreases by a factor of 5 (from 10 to 2).  However, the values of ($D$$\times$$A$)/$P$ for the best and 10th highest probability models are comparable (0.24 and 0.32). This simple comparison offers some assurance that we have not simply picked out a single model that by chance describes the observed distribution.

Along similar lines, we also consider the 10th lowest probability models in each mass bin.  These models all have vanishingly small probabilities and exhibit parameters that are drastically different from the highest probability models.  For example, the highest probability models favor relatively large bursts in the lowest mass bin, while the 10th worst model is characterized by long periods of constant SF, with bursts of amplitude 2 occurring only $\sim$ 3\% of the time.  Conversely, in higher mass bins where highest probability models are primarily characterized by constant SFHs and modest bursts, the 10th lowest probability models have typical parameters of short duration, high amplitude bursts.

An examination of full parameter space reveals a degree of degeneracy in model parameters.  Specifically, models with similar values of ($D$$\times$$A$)/$P$ also have comparable KS probabilities.  Thus, we consider the highest probability model as a representation of a set of models with similar values of ($D$$\times$$A$)/$P$.  

\subsection{Evolution of Predicted Fluxes, Masses, and SFRs}
\label{evolution}

\subsubsection{Model SFRs versus Stellar Mass and $\Sigma_{R}$}
\label{loops}

\begin{figure}[t]
\begin{center}
\epsscale{1.2}
\plotone{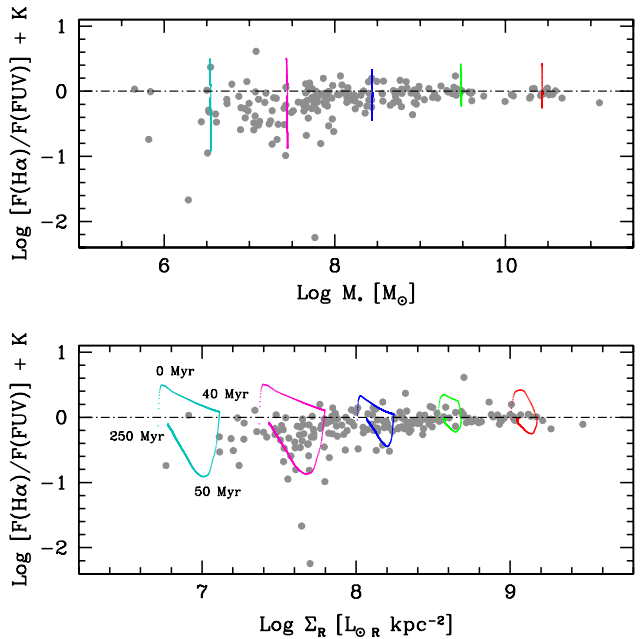}
\caption{\scriptsize {The predicted \halpha-to-FUV ratios from one burst cycle of the best fit SFHs plotted over the observational data.  The red, green, blue, magenta, and cyan color coding each refer to the best fit SFH models in the bins of highest to lowest stellar mass, as detailed in Figure \ref{best_fit_sfhs}.  To help illustrate the differences in the flux ratios, a constant $\kappa$ has been added.  The value of $\kappa$ is $-$13.17, the negative of the expected flux ratios from our models for a constant SFH.  To illustrate the evolution of the different quantities, we have plotted the models without simulated observational noise. The relative evolution in stellar mass and surface brightness of the loops has been determined entirely by the model SFHs.   In the bottom panel, the age labeling (black) corresponds to time after the burst cycle has started.  The model points are plotted for equal time intervals of 0.1 Myr.  The horizontal black dot-dashed line in each panel is the fiducial ratio.}}

\label{evol_models}
\end{center}
\end{figure}

As demonstrated in \citet{meu09}, a model SFH that matches the distribution of \halpha-to-FUV ratios may not be able to simultaneously account for other observables, such as $R$-band surface brightness or stellar mass.   With this in mind, in Figure \ref{evol_models}, we compare the evolution of the \halpha-to-FUV ratios versus stellar mass (top) and $R$-band surface brightness (bottom) over one burst cycle against the observed data. For illustrative purposes, we have excluded the simulated observational noise from this particular figure.  
 
The relative evolution of stellar mass and surface brightness have been derived entirely from the models.   We centered the starting point of the models on the median values of the observations.   This choice resulted in overlap between points from different models, and  in those cases, small adjustments ($\lesssim$ 0.3 dex) in the absolute placement of mass or luminosity were made to improve clarity.  Loops are composed of discrete model points sampled at time intervals of 0.1 Myr.  For reference, we have added time labels that correspond to distinct events in the burst cycle to the loop in the lowest surface brightness regime in Figure \ref{evol_models} (cyan points), which has a period of 250 Myr, a burst duration of 40 Myr, and an amplitude of 30 (P250, D40, A30).

\begin{deluxetable}{lccccc}
\tablecolumns{5}
\tablewidth{0pt}
\tablecaption{Model and Observed \halpha-to-FUV Ratios Relative to the Fiducial}

\tablehead{   
\colhead{Bin} &   
    \colhead{Above} &
    \colhead{Above} &
       \colhead{Below} &
       \colhead{Below}\\  
\colhead{} &
    \colhead{Observed} &
    \colhead{Model} &
       \colhead{Observed} &
       \colhead{Model} \\
       \colhead{(1)} &   
\colhead{(2)} &
    \colhead{(3)} &
    \colhead{(4)} &
       \colhead{(5)}
       }
       
\startdata    
$\log(M_{\star})$ $>$10.0 & 31.3 & 30.6 & 68.7 & 69.4 \\
10.0$\ge$$\log(M_{\star})$$>$9.0 & 35.0 & 36.5 & 65.0 & 63.5 \\
9.0$\ge$$\log(M_{\star})$$>$8.0 & 29.1 & 31.5 & 70.9 & 68.5\\
8.0$\ge$$\log(M_{\star})$$>$7.0 & 15.5 & 13.6 & 84.5 & 86.4 \\
$\log(M_{\star})$ $\le$7.0 & 21.7 & 16.3 & 78.3 & 83.7 

\enddata
\tablecomments{\small {A comparison between the observed and predicted fraction of galaxies above and below the fiducial, per stellar mass bin.  Columns (2) and (4) correspond to the observations while Columns (3) and (5) are values for the models.}}
\label{tab4}
\end{deluxetable}

During a burst, the \halpha\ and FUV SFRs are not in equilibrium.  Early in the burst, the production of \halpha\ photons exceeds that of the FUV.  However, stars that are FUV bright, are less massive, and have longer typical lifetimes than those responsible for \halpha\ production.  The difference in stellar lifetimes leads to a build up of FUV emission over the course of the burst, resulting in a net decline in the \halpha-to-FUV ratio.  We see this behavior concretely in the magenta `loop' in Figure \ref{evol_models}.  Immediately following the beginning of a burst, (denoted by `0 Myr'), the \halpha-to-FUV ratio is at its peak, as expected.  As the burst continues over the subsequent duration of 40 Myr, the \halpha-to-FUV ratio steadily declines.  
 
Following the end of the burst, we see a significant decrease in the \halpha-to-FUV ratio between 40 and 50 Myr.  There is negligible \halpha\ flux after the burst, as the more massive stars expire within a few Myr of the termination of star formation. Thus, the build up of FUV emission drives the net \halpha-to-FUV ratio to lower values.  In addition, the SFH has returned to a constant, lower level SFR during this 10 Myr period, leading to an equilibrium state for the production of  \halpha\ and FUV flux.  

Finally, from 50 to 250 Myr, we see a steady increase in the \halpha-to-FUV ratio.  The FUV excess, built up during the burst, continues to fade, and the contribution of the constant SFR to the net \halpha-to-FUV ratio increases.  At 250 Myr, near the end of the burst cycle, the \halpha-to-FUV ratio approaches the equilibrium value.  
 
We now examine the same sequence, only for the evolution of $R$-band surface brightness.  During the burst, the $R$-band surface brightness increases due to the production of luminous young stars, reaching a maximum at the end of the burst.  Relative to the initial value, the surface brightness has increased by a value of $\log(\Sigma_{R})$ $\sim$ 0.5.  For the subsequent 10 Myr, the $R$-band surface brightness decreases.  The death rate of luminous stars exceeds the birth rate, meaning the galaxy cannot sustain the maximum surface brightness level obtained during the burst phase, and consequently begins to fade.  Finally, from 50 to 250 Myr, the $R$-band surface brightness continues to fade and approaches the value consistent with that expected from a constant SFH. 

\begin{figure}[t]
\begin{center}
\epsscale{1.2}
\plotone{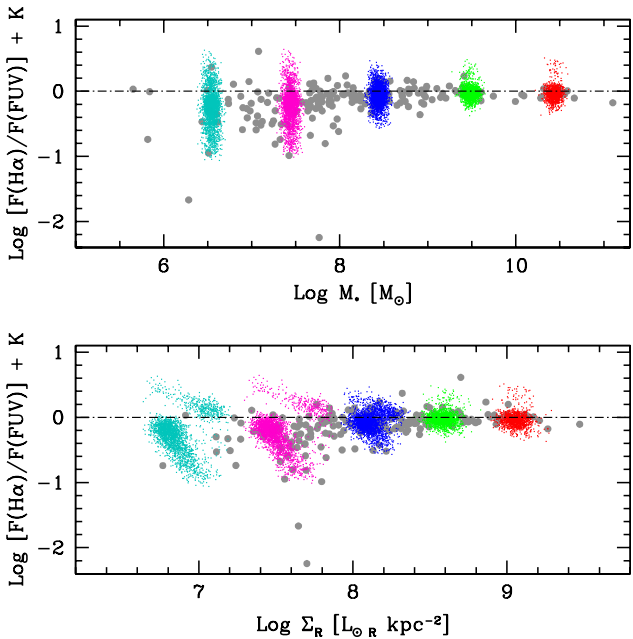}
\caption{\small {The same as Figure \ref{evol_models}, only including the effects of simulated observational noise.}}
\label{sevol_models}
\end{center}
\end{figure}

\subsubsection{Flux Evolution with Simulated Noise}
\label{noise}

In Figure \ref{sevol_models}, we now consider the model loops including the effects of simulated noise, introduced to mimic uncertainties in the observed fluxes.  The inclusion of noise adds to the dispersion in the models points.  The model points represent equal time intervals of 0.1 Myr, such that the density of points is directly proportional to the amount of time spent at a given phase of the burst cycle.

From this information, we conduct a simple comparison between the fraction of model and data points above and below the fiducial \halpha-to-FUV ratio.  As shown in Table \ref{tab4}, the predictions from the models agree with the observations within $\sim$ 5\%.  In general,  $\sim$ 30-35\% of the observed massive galaxy sample is above the fiducial. The corresponding models predict a similar fraction. Observations of lower mass systems show that $\sim$ 15-20\% of the galaxies are above the fiducial.  The highest probability SFH models in these bins also occupy a similar range, indicating that models of galaxies bursty SFHs naturally account for low mass galaxies with both low and normal \halpha-to-FUV ratios.

\subsection{\halpha-to-FUV Evolution vs. SFR(\halpha) and $M_{B}$}

\begin{figure}[t]
\begin{center}
\epsscale{1.2}
\plotone{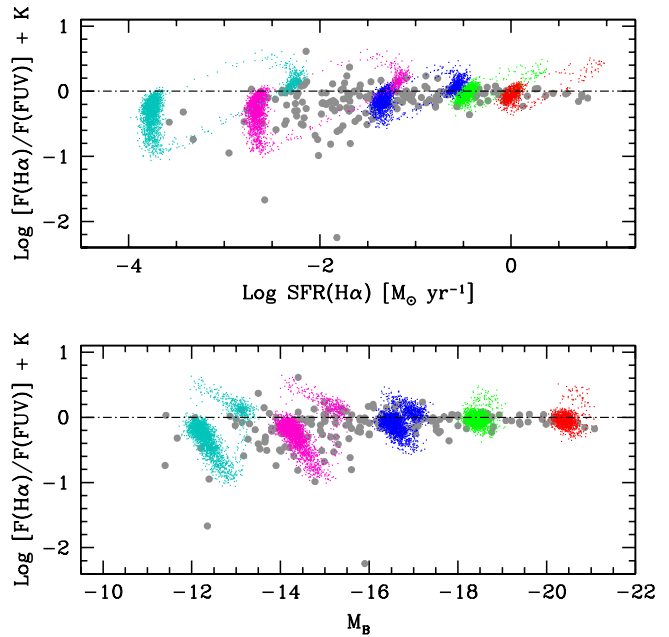}
\caption{\small{The same as Figure \ref{sevol_models}, only now plotted versus SFR(\halpha) and $M_{B}$.}}
\label{sevol_models_ha_mb}
\end{center}
\end{figure}

In Figure \ref{sevol_models_ha_mb}, we plot  the \halpha-to-FUV ratios versus SFR(\halpha) (top) and versus $M_B$ (bottom), including the effects of simulated noise as in \S \ref{noise}.  In the top panel, we see that the predicted change in the \halpha-to-FUV ratio shows an expected strong correlation with the \halpha\ based SFR, such that galaxies with higher \halpha-to-FUV ratios generally have higher \halpha\ SFRs.  From the observations, we see that galaxies with low \halpha\ SFRs and low \halpha-to-FUV ratios are quite rare, which agrees with the SFH model predictions.

Plotted versus $M_{B}$, the model SFHs indicate an increase of $\sim$ 2 mag in blue luminosity during a burst.  The majority of the time, however, the model SFHs show that galaxies are in lower luminosity states with \halpha-to-FUV ratios below the fiducial.

\section{Discussion}

\subsection{Comparison with Previous Studies of \halpha-to-FUV Ratios}
\label{previous}

Our study is not the first to model the effects of SFHs on \halpha-to-FUV ratios.  Within the past decade three notable studies \citep{igl04, bos09, meu09} have compared \halpha\ and FUV fluxes from sets of SFH models with those from observations.  These three studies represented a forward step in considering a range of SFH parameter space beyond past analysis, which were generally restricted to either constant or instantaneous burst models \citep[e.g.,][]{bua87, bua92, gla99, yan99, sul00, bel01b, moo00}.

Likewise, the analysis presented in this paper provides another incremental step toward understanding the impact of non-constant SFHs on observed \halpha-to-FUV ratios.  The primary differences between this study and those of \citet{igl04}, \citet{bos09}, and \citet{meu09} are in the construction and analysis of model SFHs and in the method of correcting for internal dust attenuation.  In this section, we will focus on the differences in model SFHs.

The model SFHs considered in this paper represent a fine sampling of SFH parameter space.  In contrast, two of the three previous studies consider a coarse sampling, generally intended to bracket ranges of SFHs.  For example, \citet{meu09} consider 18 single burst and gasp model SFHs with durations of 10, 10, 1000 Myr and amplitudes of 2, 10, and 100, and find that none are able to simultaneously account for the observed trends of \halpha-to-FUV ratios versus \halpha\ and $R$-band surface brightness.  \citet{igl04} considered a slightly broader range SFH parameters and analyze 36 model SFHs, but reach an ambiguous conclusion concerning the influence of SFHs on \halpha-to-FUV ratios.  Our SFH modeling finds that none of the extreme models, i.e., those considered bracketing the range of parameter space, provide good descriptions of the data.  Thus, it is not entirely surprising that these prior studies did not find well matched model SFHs.  

\begin{deluxetable}{lccccc}
\tablecolumns{5}
\tablewidth{0pt}
\tablecaption{Model and Observed \halpha-to-FUV Ratios Relative to the Fiducial}

\tablehead{   
\colhead{Bin} &   
    \colhead{Above} &
    \colhead{Above} &
       \colhead{Below} &
       \colhead{Below}\\  
\colhead{} &
    \colhead{Observed} &
    \colhead{Model} &
       \colhead{Observed} &
       \colhead{Model} \\
       \colhead{(1)} &   
\colhead{(2)} &
    \colhead{(3)} &
    \colhead{(4)} &
       \colhead{(5)}
       }
       
\startdata    
$\log(M_{\star})$ $>$10.0 & 31.3 & 30.6 & 68.7 & 69.4 \\
10.0$\ge$$\log(M_{\star})$$>$9.0 & 35.0 & 36.5 & 65.0 & 63.5 \\
9.0$\ge$$\log(M_{\star})$$>$8.0 & 29.1 & 31.5 & 70.9 & 68.5\\
8.0$\ge$$\log(M_{\star})$$>$7.0 & 15.5 & 13.6 & 84.5 & 86.4 \\
$\log(M_{\star})$ $\le$7.0 & 21.7 & 16.3 & 78.3 & 83.7 

\enddata
\tablecomments{\small {A comparison between the observed and predicted fraction of galaxies above and below the fiducial, per stellar mass bin.  Columns (2) and (4) correspond to the observations while Columns (3) and (5) are values for the models.}}
\label{tab4}
\end{deluxetable}

The `micro-SFHs' considered by \citet{bos09} represent a limited, but fine sampling of SFH parameter space.  The focus of micro-SFHs is on the effects of the more recent burst on the \halpha\ luminosity.  Episodic bursts with durations of 2, 5, and 10 Myr over the course of 100 Myr were considered to simulate the effects of various \hii\ regions turning on and off.  The principle behind this method is similar to our analysis, except we consider a larger number of possible model permutations and the contribution of ancient stellar populations, allowing us to accurately assess the impact of recent SFHs on stellar mass and surface brightness, as well as the observed \halpha\ and FUV fluxes.  We further note that the models considered by \citet{bos09} are similar to those in our highest mass bins, but they do not consider models with parameters such as those in our lowest mass bins.  This seems appropriate, however, as the sample of \citet{bos09} does not extend to comparably low luminosity galaxies.

A further point of contrast between our study and the previous three is the division of the sample for analysis.  Each of the previous studies attempt to find a single SFH that best describes the entire observed sample.  By binning the data in increments of stellar mass, we do not impose the restriction that a single model SFH must be able to explain the observed \halpha-to-FUV ratios in both massive galaxies such as M51 and low mass dwarfs such as GR8.  Indeed, as we have shown, the highest probability models converge on drastically different characteristic SFHs for galaxies of different masses.

\subsection{SFHs and the IMF}

An underlying assumption to any calibration of SFR indicators is the nature of the IMF.  In the conversion from observed flux to SFR, the IMF is assumed to be universal, i.e., the same with respect to time and environment, and fully populated, i.e., the relative number of stars per unit mass interval is always the same.   However, whether either of these criteria are true in nature remains a heavily debated and open question (see the review by \citealt{bas10}).  

Unfortunately, only a handful of variable IMF scenarios in the literature have been discussed in the context of \halpha-to-FUV ratios. Extracting results from some of these studies, we proceed to discuss these findings alongside predictions from the model SFHs analyzed in this paper.

\subsubsection{Stochastically Sampling the IMF}

We first consider the case of a universal, but not fully sampled IMF.  The effect of this stochastically sampled IMF scenario on the \halpha-to-FUV ratio has been simulated in \citet{lee09b}.  Assuming a constant SFH, the authors found that the predicted turnover in \halpha-to-FUV ratios occurred at too low an \halpha\ luminosity relative to the data (see Figure 7 in \citealt{lee09b}). That is, randomly sampling a universal IMF does not appear to adequately account for the observations.  

This result is confirmed by more recent simulations in \citet{das11}, \citet{fum11}, and \citet{eld11}.  In each paper, the authors verify the inability of random sampling of the IMF to account for the observed trend in \halpha-to-FUV ratios, under the assumption of a constant SFH and a fully populated cluster mass function.  Instead, each study demonstrates that stochastic sampling of \emph{both} the stellar and cluster mass functions are necessary to produce \halpha-to-FUV ratios that are consistent with observations.  The effect of stochastically sampling the cluster mass function is that lower mass galaxies form massive clusters less frequently than higher mass systems. Along the same lines, stochastically sampling stellar IMF implies that high mass stars have a higher probability of forming in high mass clusters.  In light of these scenarios, we would then expect low mass galaxies to form fewer high mass clusters, and hence fewer high mass stars when compared to more massive galaxies.  Therefore, low mass systems would have fewer massive stars capable of producing \halpha, and the \halpha-to-FUV ratios in low mass systems is expected to be less than for more massive galaxies.  Crucially, these models posit that the formation of massive clusters in low mass galaxies are rare, but not impossible, which is the case for certain prescriptions of a variable IMF (see \S \ref{igimf}).  More detailed discussion and quantitative analysis of these effects can be found in \citet{das11}, \citet{fum11}, and \citet{eld11}.

\subsubsection{SFHs and the Integrated Galactic IMF}
\label{igimf}

The second possibility we consider is that the IMF is not universal.  This scenario has been widely debated in the literature, but a consensus has yet to be reached \citep[e.g.,][]{bas10}.  We focus on one particular scenario, the Integrated Galactic IMF (IGIMF) as detailed in \citet{wei05}, which makes specific predictions for \halpha-to-FUV ratios as functions of \halpha\ luminosity \citep[][]{pfl07, pfl09}.

There are three tenets of the IGIMF: $(i)$ all stars are formed in clusters; $(ii)$ the maximum mass of a cluster star is determined by the mass of the cluster in which it is formed; $(iii)$ the maximum cluster mass is a deterministic function of the integrated SFR of a galaxy.   From these three postulates, the IGIMF predicts a systematic increase in the ratio of \halpha-to-FUV fluxes as a function of \halpha\ luminosity, under the assumption of a constant underlying SFH \citep[][]{pfl07, pfl09}.  

In what follows, we will compare predictions from the IGIMF and model SFHs with our observations.  First, we will consider the standard case of the \halpha-to-FUV ratios as a function of \halpha\ luminosity.  The IGIMF predictions in this case are taken directly from the work of \citet{pfl09}.  We will then compare the model predictions to the data as a function of galaxy stellar mass.  In this case, to compute the galaxy stellar mass for the IGIMF models, we have assumed a constant SFH over the history of the galaxy at a level of the current \halpha\ SFR.  We discuss on the validity of this assumption below.

\begin{figure}[t]
\begin{center}
\epsscale{1.2}
\plotone{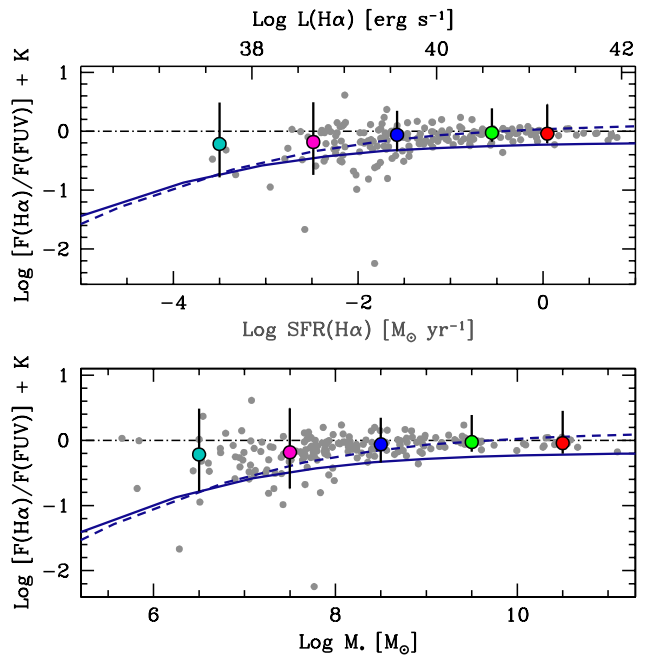}
\caption{\scriptsize{The observed \halpha-to-FUV flux ratios (grey points) versus \halpha\ luminosity (top panel) and stellar mass (bottom panel).  The large colored point correspond to the median values from the best fit SFH modes as described in \S \ref{results}.  The error bars on the models correspond to the 5th and 95th percentiles of the \halpha-to-FUV flux distributions for each model SFH.  The dashed and solid navy lines correspond to predictions of the IGIMF `minimal1' and 'standard' models, respectively, as described in \citet{pfl07, pfl09}.  The black dot-dashed line represents the fiducial ratio for our models.  Axes represented \halpha\ and FUV SFRs are grey as they are not valid in regions of low SFR(\halpha) according to the IGIMF models.  The two model types are indistinguishable in the upper plot when the flux ratios is considered versus \halpha\ luminosity.  However, when plotted versus galaxy stellar mass, the IGIMF models are no longer consistent with the majority of points at low masses, which lie above the model curves.}}
\label{igimf_comp}
\end{center}
\end{figure}

As a first comparison, we follow the discussion of the IGIMF in \citet{lee09b}, and  over-plot predictions for the  `standard' and `minimum1' IGIMF models (solid and dashed navy lines) on the observed \halpha-to-FUV ratios versus \halpha\ luminosity in the top panel of Figure \ref{igimf_comp}.  The two IGIMF models correspond to different convolutions of the cluster mass function with a stellar IMF, such that the minimal1 model is the least deviant from a standard Salpeter IMF (see \citealt{pfl07, pfl09} for specific details).  We have adopted the IGIMF models from \citet{pfl09} that assume a constant SFH over the history of the universe.
 
We have also over-plotted the predictions from the highest probability model SFHs.  Each of the colored points represents the median \halpha-to-FUV ratio from models shown in Figure \ref{best_fit_sfhs} and listed in Table \ref{tab3}.  The placement on the horizontal axis corresponds to the median \halpha\ luminosity in each mass bin for the observed sample.  The error bars represent the 5th and 95th percentiles ($\sim$ 2-$\sigma$) in the model distributions.  

In the top panel of Figure \ref{igimf_comp}, we see that both the IGIMF and SFH models qualitatively agree with observations.  At high \halpha\ luminosities, both types of model indicate that ratios should be near the fiducial. For decreasing \halpha\ luminosity, both models predict a decline in the \halpha-to-FUV ratio that is in reasonable agreement with the mean observational trends.  Of the two IGIMF scenarios, the `minimal1' model appears to more closely follow the mean of the data better than the `standard1' model, particularly when the data are considered as a function of stellar mass.  

The scatter in the \halpha-to-FUV ratios is another quantity we consider. For \halpha\ luminosities $\gtrsim$ 10$^{38.5}$, we see a number of data points that lie within the range of the SFH model predictions, but above the two IGIMF curves.  We note that it is not clear how the IGIMF can explain galaxies that lie significantly above the predicted curves, and pursue a more detailed discussion of this point below.

In the bottom panel of Figure \ref{igimf_comp}, we repeat the comparison outlined above, only this time we plot galaxy stellar mass on the x-axis.  Again, we see that both the SFH and IGIMF models predict a general decline in the \halpha-to-FUV ratios as a function of decreasing stellar mass.  However, both IGIMF models tend to under predict the \halpha-to-FUV ratios for galaxies $\lesssim$ 10$^{8}$ \msun. In comparison, the median values of the SFH models do not decrease as sharply as those of the IGIMF, and instead exhibit a larger dynamic range of possible model values.

Of particular interest is the relationship between the two types of models and the scatter in the data.  The broad dispersion of \halpha-to-FUV ratios predicted by the model SFHs are generally in good agreement with the observations.  The most notable discrepancies are the handful of galaxies that have \halpha-to-FUV ratios that are lower than predicted by the models.  This may be an indication that stochastic IMF effects need to be included to explain the extremely low \halpha-to-FUV points.  

The IGIMF models are generally not in good agreement with the observed scatter.  In the best case, the IGIMF models track the decline of average \halpha-to-FUV ratios as a function of decreasing \halpha\ luminosity, but there are a number of galaxies with ratios significantly below the IGIMF predicted curves, and a smaller fraction that lie above.  The disagreement worsens when the comparison is done as a function of stellar mass.  In particular, for galaxies with stellar masses $\lesssim$ 10$^{8}$ \msun, there are a substantial number of galaxies that have \halpha-to-FUV ratios significantly above the IGIMF predictions.  

These data points are challenging to understand in the framework of the IGIMF.  Taken at face value, the IGIMF is a deterministic theory \citep[e.g.,][]{pfl09}.  That is, it makes specific predictions about the the functional form of the \halpha-to-FUV ratio, namely that \halpha\ should be deficient in galaxies with low masses or low SFRs. Yet, the data show a number of low mass systems with normal or high \halpha-to-FUV ratios, implying the frequent presence of massive star(s) in low mass systems; a observational finding that does not seem compatible with the tenets of the IGIMF.  

Admittedly, there are two areas of uncertainty with this interpretation.  First, the connection between total galaxy stellar mass and the IGIMF is not well defined in the literature.  Although such a connection has been invoked in other analyses of the IGIMF \citep[e.g.,][]{wei05, elm06, bos09}, extrapolating the current \halpha\ SFR to a constant lifetime SFH to determine the stellar mass is not a valid assumption.  In the event that we have \emph{over-estimated} the galaxy stellar mass, the IGIMF curves would shift to the left in the bottom panel of Figure \ref{igimf_comp}.  In order to match the mean observed flux ratios, the extrapolated galaxy masses need to be decreased by a factor of $\sim$ 10.   Even with this correction applied, the IGIMF curves still do not match the observed scatter in the data very well.

The second area of uncertainty is the connection between the IGIMF and assumed form of the SFH.  Both the `standard' and `minimal1' IGIMF  models assume a constant SFH \citep[e.g.,][]{pfl07, pfl09}.  Including the effects of a bursty SFH may be able to produce \halpha-to-FUV ratios higher than currently predicted.  As seen from the model SFHs, during the early phases of a burst, the \halpha-to-FUV ratios are near their peak.  Combining this mechanism with the IGIMF provides at least one possible explanation for galaxies with higher than predicted \halpha-to-FUV ratios.

However, the convolution of bursts with the IGIMF is not without complications.  First, if bursts are required to explain \halpha-to-FUV ratios above the IGIMF predictions, this implies that all such galaxies are currently in the midst of a burst.  However, it seems unlikely that the majority of low mass galaxies in our sample should be undergoing simultaneous bursts.  Second, under the auspices of the IGIMF, one also needs to consider the absolute SFR, not just the relative burst amplitudes.  At low absolute SFRs such as 10$^{-4}$ \msun\ yr$^{-1}$, the IGIMF model predict that $\log[F(H_{\alpha})/F(FUV)]$ is $\sim$ $-$1.  Extrapolating this SFR over 13.7 Gyr yields a stellar mass of $\sim$ 10$^{6}$ \msun.  At this stellar mass, we see a handful of galaxies with $\log[F(H_{\alpha})/F(FUV)]$ $\gtrsim$ 0.  The only way for such low mass galaxies to have these high \halpha-to-FUV ratios is through intense bursts of SF that significantly increase the amount of \halpha\ emission.    Additionally, we note that such bursts should not be too long in duration or too high in frequency or the final mass of the galaxy  could be substantially increased.  

Overall, we find that bursty SFH models can qualitatively account for the general decline and scatter in the observations as a function of decreasing \halpha\ luminosity or galaxy stellar mass, whereas the simplest form of the IGIMF models do not.  We emphasize that this finding does not invalidate the IGIMF models, nor suggest that a variable IMF is not possible, only that there appear to be other viable explanations for the observed \halpha-to-FUV ratios that do not involve a systematically varying IMF. Indeed, we believe that future model efforts that include more complex SFHs, and reasonable variations on the IMF are essential for understanding SF processes in galaxies.  We suggest some possible paths forward in the following section.

\subsection{Toward More Realistic SFH Models}
\label{real_sfhs}

In this paper, we have found a set of SFH models that match the trend and scatter in observed \halpha-to FUV ratios.  However, we caution that the broad applicability of our conclusions should be tempered by the inherent simplicity of the models.  SFHs in nature are undoubtably more complex than periodic bursts with fixed amplitudes and durations.  We have considered characteristic SFH models for galaxies in comparable mass ranges, but it is likely that even galaxies with similar stellar masses do not share \emph{identical} SFHs.  Indeed, studies within the Local Group and nearby universe have shown that galaxies with similar masses exhibit diversity in their SFHs \citep[e.g.,][]{dol05, wei11}.

Determining the degree to which any model SFH truly reflects those in nature is a challenging and presently unresolved question.  A comprehensive assessment of the realistic nature of the model SFHs used in this paper is beyond the scope of this study, but is the subject of future work \citep[e.g.,][]{johnprep, weiprep}.  Instead, in this section we outline the limitations of the simple model SFHs and discuss ways in which future SFH model efforts can be improved.

One increasingly common method of directly measuring SFHs is from analysis of color-magnitude diagrams (CMDs) of resolved stellar populations in nearby galaxies (see \citealt{tol09} and references therein).  In Figure \ref{real_sfhs}, we show an example of six such SFHs measured from CMDs based on Hubble Space Telescope imaging of dwarf galaxies in the M81 Group \citep{wei08, dal09, wei11}.  The solid black and dot-dashed magenta lines represent the same SFH binned to different time resolutions, which we discuss below.

  These measured SFHs provide an indication of the complexity of galaxy-wide star formation.  The most striking characteristic in these measured SFHs is the degree of stochasticity in the amplitude and duration of star formation episodes.  Further, there is no clearly defined period associated with any of the measured SFHs.  There is also a conspicuous lack of stochasticity in the model SFHs (e.g., Figure \ref{sample_sfhs}).  Clearly, non-uniform parameterization of bursty SFHs is an important next step toward constructing more realistic SFHs.

Of course, there is some degree of similarity between the modeled and measured SFHs.  Focusing on the most recent 100 Myr (insets in Figure \ref{real_sfhs}), where the time resolution of the CMD-based SFHs is the highest, a handful of bursts with amplitudes of $\sim$ 10-50 are present in the measured SFHs, similar to those found in the highest probability models for low mass galaxies.  The measured SFHs also suggest that extended duration, high amplitude bursts (e.g., $D \gtrsim$ 100 Myr, $A \gtrsim$ 50) are probably not common.  Burst durations of hundreds of Myrs have been measured in a minority of starburst dwarf galaxies \citep{mcq10a,mcq10b}, but with average burst amplitudes of less than a factor of 5.  However, these global, long duration bursts are superimposed with a short timescale `flickering'  of enhanced SF, which result in short-term burst amplitudes $\gtrsim$ 10 on 10 Myr timescales.

\begin{figure}[t]
\begin{center}
\epsscale{1.2}
\plotone{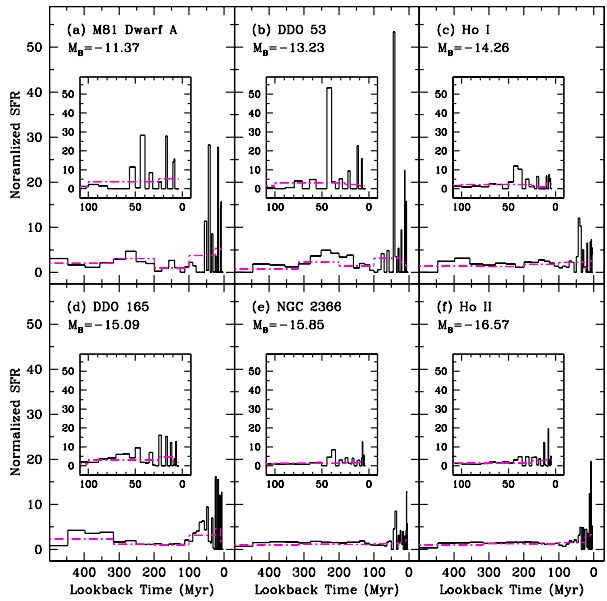}
\caption{\scriptsize{Select SFHs measuredf from analysis of HST-based optical CMDs \citep{wei08}.  The SFHs plotted as black solid lines are shown at the highest possible time resolution, while the dot-dashed magenta lines indicate the more conservative time binning scheme adopted in \citet{wei08}.  The insets show the most recent 100 Myr of each SFH, where the CMDs provide for high time resolution on the derived SFHs.  Errors on the SFRs have been omitted for clarity, but the increase in time bin size generally corresponds to smaller uncertainties in the SFRs.  We draw attention to two features: the amplitudes of SF events and the complexity of the measured SFHs.  Characteristic amplitudes in the measured SFHs are qualitatively similar to those in the best fit model SFHs, but clearly a more precise analysis is needed to quantify the similarity.  Relative to the simple model SFHs we consider, the measured SFHs show stochastic variations in both the frequency and duration of SF events, as well as their amplitudes.  Incorporating such features into future SFH models in an important step toward making them more realistic.}}
\label{real_sfhs}
\end{center}
\end{figure}

Average burst amplitudes in these systems are typically a factor of a $\lesssim$ 5 over the course of hundreds of Myr, although variations in the SFRs on timescales of order $\sim$ 10 Myr can vary by factors of 10 or more.   Determinations of  exact amplitude and duration parameters directly from CMD-based SFHs are challenging due to the logarithmic time resolution of the measured SFHs. We explicitly illustrate the issue at hand in Appendix \ref{timeres}.

Integrated fluxes provide further empirical constraints for SFH modeling efforts.  A realistic set of model SFHs should be able to reproduce such observations as the distribution of observed \halpha\ EWs, the UV luminosity function, and optical broadband colors and luminosities.  The simple model SFHs we consider are not in strong agreement with either observed distributions of either \halpha\ EW or UV luminosity, as presented in \citet{lee09a} and \citet{lee11}.  While both the model and observed distributions are well described by lognormal functions, the functional form of the model predictions has too many galaxies near the median value and not enough at the extremes.  Some of the difference can be reconciled by increasing the simulated noise in the models, but we would prefer to avoid this type of fine-tuning.  Further, the addition of stochasticity would also mitigate the over-density of galaxies near the median of the distribution.  Quantitatively testing these effects is beyond the scope of this paper, but is discussed in the context of the SLUG population synthesis code \citep[][]{fum11, das11}.

\section{Summary}
\label{conclude}

We have compared predictions from simple, periodic SFHs with observed \halpha-to-FUV SFR ratios observed as part of the 11HUGS and \emph{Spitzer} LVL programs. From a suite of $\sim$ 1500 simple model SFHs, we have identified a set of models that matches the observed distribution of \halpha-to-FUV SFR ratios over a range of galaxy stellar masses. We find that high mass galaxies tend to have characteristic SFHs that are predominantly constant, with relatively modest amplitude bursts ($A$ $\lesssim$ 10) spaced by $\sim$ 100 Myr. We find that lower mass galaxies are best described by SFHs with burst amplitudes of $\sim$ 30, and inter-burst spacings of $\sim$ 150-200 Myr.   These model SFHs have burst amplitudes that are comparable to those inferred from CMDs of nearby star forming dwarf galaxies.

Using results from the highest probability models, we compared the predicted evolution of the \halpha-to-FUV ratios versus stellar mass, $R$-band surface brightness, SFR(\halpha), and $M_{B}$, and found that in all cases, the models were in good agreement with observations.

We find that highest probability models are well matched to the observed systematic decline in the \halpha-to-FUV ratios as a function of both decreasing \halpha\ luminosity and decreasing galaxy stellar mass.  

Variations in the high mass stellar IMF have been proposed as an explanation for this trend.  We therefore compare predictions from the model SFHs with those from the IGIMF theory, which also makes specific predictions for \halpha-to-FUV ratios.  From this  comparison we see that stochastic sampling of the IMF cannot re-produce the observed trend.  A convolution of stochastically sampled stellar and cluster mass functions are necessary to match observations. We find that both the SFH and IGIMF predictions are in good agreement with the \halpha-to-FUV ratios as a function of \halpha\ luminosity.  However, when considered versus stellar mass, it is not clear if the IGIMF allow for \halpha-to-FUV ratios higher than the predicted curves, whereas the model SFHs naturally account for the observed scatter.  

Instead, the we find that only time variable SFHs are able to explain both the mean trend in the data, as well as the scatter.  While we have only considered toy model SFHs, they share characteristics with the stochastically sampled cluster and stellar mass function scenario presented in \citet{das11}, \citet{fum11}, and \citet{eld11}.  Essentially, at low SFRs, a stochastically sampled cluster mass function rarely produces massive clusters.  However, when a massive cluster does form, the SFR of the galaxy increases, and can effectively look like a burst of SFH.  While the frequency of massive cluster formation in low mass galaxy is not well understood, the new population synthesis code, SLUG \citep{das11, fum11} will permit investigations on the impact of stochastic cluster formation on the galaxy wide SFH and broadband flux properties.

Finally, we discuss the limitations of our toy model SFHs, including the assumptions of periodicity and lack of stochasticity.  In addition, we find that one major challenge is understanding differences between our model predictions and observations of the \halpha\ and UV luminosity functions, as well as precisely matching observed optical and NIR galaxy luminosities.  Exploring these differences are one of the topics that will be explored in the upcoming paper of \citet{johnprep}.

\acknowledgments

The authors would like to thank the anonymous referee for useful suggestions that helped to improve this paper.  The authors would also like to thank Nate Bastian, John Eldridge, and Robert da Silva for insightful discussions about the nature of the IMF.  DRW is grateful for support from the University of Minnesota Doctoral Dissertation Fellowship and Penrose Fellowship. This work is based on observations made with the NASA/ESA Hubble Space Telescope, obtained from the data archive at the Space Telescope Science Institute.  Support for this work was provided by NASA through grants GO-10605 and GO-10915 from the Space Telescope Science Institute, which is operated by AURA, Inc., under NASA contract NAS5-26555. This research has made use of the NASA/IPAC Extragalactic Database (NED), which is operated by the Jet Propulsion Laboratory, California Institute of Technology, under contract with the National Aeronautics and Space Administration.  This research has made extensive use of NASA's Astrophysics Data System Bibliographic Services.

\clearpage

\begin{appendix}
\section{The Effects of Logarithmic Time Resolution of CMD-based SFHs for Deriving Model SFH Parameters}
\label{timeres}

In principle, one can synthesize \halpha-to-FUV ratios from CMD based SFHs, eliminating the need for the intermediate step of modeling SFHs.  However, the non-uniformly linear time resolution of CMD based SFHs presents a challenge to such efforts.  Much like stellar isochrones, CMD based SFHs are naturally sampled in uniform logarithmic time steps, which are not equally spaced in linear time.  The combined coarseness of the time binning for lookback times $\gtrsim$ 100 Myr and the sensitivity of the \halpha-to-FUV ratios to the input parameters limit the utility of CMD based SFHs for this particular purpose.  

\begin{figure}[bh]
\begin{center}
\epsscale{0.8}
\plotone{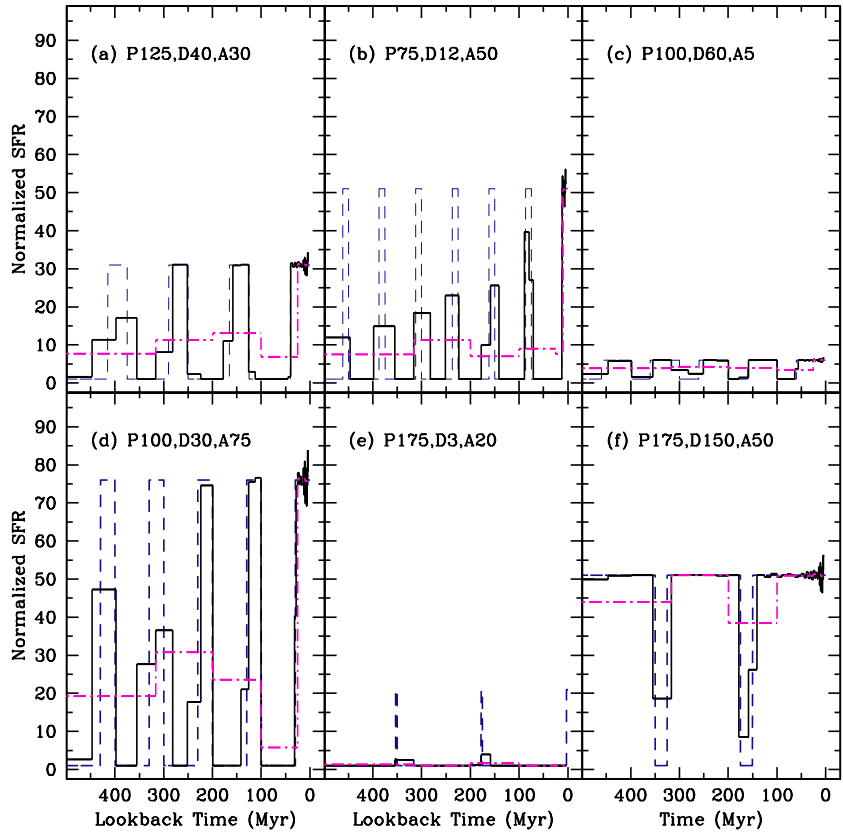}
\caption{\small{The sample model SFHs presented in Figure \ref{sample_sfhs}, resampled at the highest time resolution binning available for CMD based SFHs (solid black lines) and the broader binning scheme of \citet{wei08} in the magenta dot-dashed line.  For reference, the original model SFHs have been plotted in the navy dashed lines. The native logarithmic time binning scheme for CMD based SFHs can make it difficult to differentiate between various SFH models for times $\gtrsim$ 100 Myr.  Bursts that are closely spaced, low in amplitude, or short in duration tend to get washed out and appear as relatively small deviations above an otherwise seemingly constant SFH.  In addition to challenges posed by the logarithmic nature of CMD-based SFHs, there are degeneracies in certain stellar features, e.g., the luminous MS, and observational effects such as differential extinction which can further reduce the time resolution.  See \citet{mcq10a} for further discussion of CMD time resolution effects.}}
\label{res_sfhs}
\end{center}
\end{figure}

As a simple demonstration of the effects of time resolution, in Figure \ref{real_sfhs}, we show the same SFHs binned at the highest time resolution (solid black) and the resolution adopted by \citet[][]{wei08} (dot-dashed magenta).  Although these represent the same underlying SFH, the choice in time-binning gives a strikingly different impression of the galaxy SFHs.  Extending this exercise to the models considered in this paper, in Figure \ref{res_sfhs} we plot the example model SFHs from Figure \ref{sample_sfhs} convolved to the two CMD based time binning schemes.  In most cases, the input SFHs are sufficiently degraded to an unrecognizable state.  There is, of course, some dependence on the particular SFH, as well as the quality of the observed CMD.  A more detailed discussion of time resolution and SFH recovery from CMD can be found in \citet{mcq10a}.

\end{appendix}

\end{document}